\documentclass[prd,showpacs,preprintnumbers,nofootinbib,aps,twocolumn]{revtex4}
\usepackage{subeqnarray}
\usepackage{epsfig,amsmath,amssymb}
\usepackage{mathrsfs}
\usepackage[usenames,dvipsnames]{color}
\usepackage[pagebackref=true, colorlinks=true]{hyperref}
\definecolor{redish}{rgb}{0.7,0.2,0.0}  
\definecolor{bluish}{rgb}{0.2,0.5,0.8}
\hypersetup{linkcolor=redish,          
                  citecolor=blue,        
                  filecolor=magenta,      
                  urlcolor=bluish}          

\DeclareFontFamily{U}{rsfs}{}         
\DeclareFontShape{U}{rsfs}{m}{n}{<5> rsfs5 <6><7> rsfs7          %
  <8><9><10><10.95><12><14.4><17.28><20.74><24.88> rsfs10}{}     %
\DeclareMathAlphabet{\mathfs}{U}{rsfs}{m}{n}                     %

\def \({\left(}
\def \){\right)}
\def \[{\left[}
\def \]{\right]}

\def\pb#1{\rlap{\lower1.5ex\hbox{$\longleftarrow$}}{#1}}
\def\dpb#1{\rlap{\lower1.5ex\hbox{$\Longleftarrow$}}{#1}}
\def\spb#1{\rlap{\lower1.5ex\hbox{$\leftarrow$}}{#1}}
\def\sdpb#1{\rlap{\lower1.5ex\hbox{$\Leftarrow$}}{#1}}
\begin{document}
\title{Variation of rest mass scale in a gravitational field}
\author{Abhishek Majhi}%
\email{abhishek.majhi@gmail.com}
\affiliation{Indian Statistical Institute,\\Plot No. 203, Barrackpore,  Trunk Road,\\ Baranagar, Kolkata 700108, West Bengal, India\\}
\begin{abstract}
	I argue that an angular momentum scale is necessary to explain energy-momentum propagation along a single null geodesic, the scale being known as Planck's constant $(h)$.  If $h$ and  $c$ (the velocity of light in vacuum), are considered to be given fundamental (constant) scales in local measurements, then the rest mass scale varies exactly inversely as the time scale varies  in two different locally flat regions of a curved spacetime. As the time scale variation gives rise to gravitational redshift, the rest mass scale variation  leads to a  change in the  Compton length scale associated with an elementary particle. I suggest an experiment for the MICROSCOPE satellite and report the expected outcome. 

\end{abstract}
\maketitle
{\it Introduction}: 
Three most important physical dimensions that one perceives of in daily life  are `mass' $(M)$, `length'$(L)$ and `time'$(T)$, which are measured with {\it predefined} reference scales (or units) such as kilogram, metre and second, respectively \cite{standards,scale}. In special relativity, Einstein showed that the notion of length scale and time (or frequency) scale are not absolute concepts and vary in two frames with constant relative velocity (inertial) \cite{einsr}. What is fundamental in special relativity is the velocity scale denoted by $c$, which is the `velocity' of light in vacuum \cite{einsr}. Another scale that is absolute, but not fundamental, in special relativity is the rest mass scale $m_0$ associated with a particle \cite{restjust}. It is  `absolute' in the sense that it is invariant under Lorentz transformations i.e. $p^Ip_I=-m_0^2c^2$ where $p^I$ is the four-momentum \cite{landau, jackson}. It is not `fundamental' because it is arbitrary. 
$m_0$ represents the rest mass scale of an {\it elementary} particle, which is rather  understood as the Lorentz invariant mass scale associated with a field (e.g. see \cite{peskin}). 

 Now, to address the effect of gravitation on light propagation, in ref. \cite{einequi}, just by assuming  equivalence principle \cite{equivalence} and considering mass-energy conservation principle,  Einstein concluded that the energy scale differs in two frames that differ by a  gravitational potential due to a homogeneous field or equivalently in  constant relative acceleration (non-inertial). However, while writing down the formula for gravitational redshift, Einstein simply associated `frequency' with `energy' and therefore, implicitly assumed the existence of  a fundamental angular momentum scale that remains unaffected by gravity.  The experimental verification of gravitational redshift by Pound and Rebka  \cite{poureb1,poureb2} shows that this fundamental scale is none other than the Planck's constant $(h)$. In the language of curved spacetime geometry, Einstein's result is a manifestation of energy scale variation between two locally flat spacetime regions when the local observations are compared in a suitable coordinate system such that the Newtonian potential, with some approximation, implies the curvature \cite{wald}. However, the role of $h$ remains indispensable. For example, one can see section (6.3) of ref. \cite{wald}, where the author `makes sense' of the calculation by relating `frequency' with  `energy' and requiring  $h$ in the process (and hence invoking `quantum' mechanical concept of `photon' in the context of `classical' gravity).  I argue that     a fundamental scale of angular momentum dimension, which is none other than $h$, is necessary to make sense of energy-momentum propagation along a {\it single} null-geodesic. It has, {\it a priori}, nothing to do with quantum mechanics and this is in sharp contrast with the presently accepted general understanding.

Now, considering $h$ and  $c$ as fundamental scales, `kilogram' is defined in terms of $\Delta\nu_{cs}$,  which is the frequency scale associated with the transition between the hyperfine states of the unperturbed ground level of a Caesium (133)  atom,  in its rest frame \cite{standards}. I argue that if one considers $h$ and $c$ to be {\it given}, rather than determined, fundamental scales in local measurements, then the rest mass scale associated with an elementary particle varies exactly inversely as the time scale varies in two different locally flat regions of a curved spacetime. This is a hitherto unexplored, or rather ignored, facet of Einstein's equivalence principle \cite{caution}.  Nonetheless, the effect is quite clear from the definition of `kilogram' in terms of $\Delta\nu_{cs}$. Since frequency scale variation suggests that $\Delta\nu_{cs}$ have different meanings at different locally flat regions of a curved spacetime \cite{bl}, so does the notion of `kilogram'. Irrespective of this fact,  I suggest a very doable experiment to directly test the rest mass scale variation, along with a predicted outcome for the MICROSCOPE satellite mission \cite{microtest}. 

  {\it A review of Einstein's thought process:}  With an aim to investigate the role of $h$ in the explanation of energy propagation along a null geodesic and  to shed light on the  mass scale variation due to gravity, I  
 briefly mention the relevant results which were reported by Einstein in ref. \cite{einequi}. Einstein considered the following thought process: 
 \begin{enumerate}
\item  	Two observers $O_1$ and $O_2$, relatively at rest with respect to each other, are  equipped with a set of {\it identical} measuring instruments (implies identical scales for measurements).

\item Now, these observers, along with their corresponding set of instruments,  sit at different potentials of a homogeneous gravitational field or {\it equivalently} have a relative non-zero acceleration (equivalence principle).

\item A certain amount of energy $E_1$, as measured by $O_1$ locally in its rest frame, is emitted, in the form of radiation, towards $O_2$. 

\item $O_2$ receives this energy and  performs a local measurement in its rest frame to yield the result $E_2$.
 \end{enumerate}

     Einstein showed that, if energy conservation principle holds, then, approximately up to the first order in $\Phi/c^2$, $E_1$ and $E_2$ are related by the following equation  
\begin{eqnarray}
E_2=E_1\(1+\frac{\Phi}{c^2}\)\label{e1e2}
\end{eqnarray}
where $\Phi$ is the potential difference between $O_1$ and $O_2$. Then, Einstein wrote the formula for gravitational redshift. To do that, Einstein simply assigned `frequency' corresponding to radiation energy (and {\it not} `intensity'): {\it ``If the radiation emitted ...... had the frequency....''} (see the very beginning of section 3 of \cite{einequi}). He arrived at the formula 
\begin{eqnarray}
\nu_2=\nu_1\(1+\frac{\Phi}{c^2}\)\label{nu1nu2}.
\end{eqnarray}
Einstein's  argument to interpret the frequency shift was based on the fact that the time scale (unit) for $O_1$ and that for $O_2$ do not remain identical due to the difference in gravitational potential.

{\it Einstein's untold assumption:} It is interesting  to note that Einstein did not use or mention, {\it explicitly}, the involvement of any fundamental    scale of angular momentum dimension, that remains unaffected by gravity \cite{interesting}. However,  without this assumption  it is impossible to pass on from eq.(\ref{e1e2}) to eq.(\ref{nu1nu2}). The experimental verification of Pound and Rebka \cite{poureb1,poureb2}, indeed relied on methods, namely Mossbauer effect \cite{moss}, that involve $h$. Therefore, Einstein's {\it untold assumption} was that, besides $c$, $h$ is another fundamental scale that remains identical for $O_1$ and $O_2$ while performing local measurements. In what follows, I explain the necessity of this assumption.

{\it An angular momentum scale and null geodesic:} In special relativity, for a massive {\it point}\cite{point} particle following timelike geodesic with four velocity $u^{\mu}$, one has $
u^\mu u_\mu=-1$ and for four momentum $p^\mu=m_0 u^\mu$ one has $p^\mu p_\mu=-m_0^2c^2$ which implies
\begin{eqnarray}
&& -E^2/c^2+|\vec p|^2=-m_0^2c^2
\end{eqnarray}
where $p^0=E/c$, $m_0$ is the rest mass scale uniquely characterizing the particle under consideration and $\vec{p}$ is the spatial momentum with respect to the observer. All these quantities are dimensionally well defined and the physical interpretation of the energy carried by the massive particle along the time-like geodesic is well posed in terms of rest mass energy (Lorentz invariant) and kinetic energy (observer dependent). The role of $m_0$ becomes justified while one checks that in the non-relativistic limit it plays the usual role in the kinetic energy term $\frac{1}{2}m_0v^2$ (see p.27 of  ref. \cite{landau}). 

To seek a similar understanding about the null geodesic is perfectly legitimate. However, for propagation along a null geodesic, there is no associated rest mass scale i.e. $p^\mu p_\mu=0$, which implies
\begin{eqnarray}
&& E^2/c^2=|\vec p|^2.\label{lightenergy}
\end{eqnarray}
There is no expression for $E$ or $|\vec p|$ that can be cooked up from the Lorentz invariant energy-momentum tensor of electrodynamics. This is because, eq.(\ref{lightenergy}) is a statement about propagation along a single ray at the velocity of light (with respect to any observer), which is a {\it single} null geodesic in the spacetime description. Whereas calculation of energy-momentum tensor involves an extended region of spacetime i.e. a bundle of rays or a congruence of geodesics are required and hence, the components of the energy-momentum tensor are flux densities \cite{landau,bornwolf}. Further, added to all the above explanations, there is no non-relativistic limit of the null geodesic scenario from where one can draw any conclusion.

The only meaningful quantity that one can consider along a single null geodesic is the phase of the propagating light which is a Lorentz invariant quantity in special relativity \cite{jackson}:
\begin{eqnarray}
	\frac{(x-ct)}{\lambda}=\frac{(x'-ct')}{\lambda'}
\end{eqnarray}
where $(x,t)$ and $(x',t')$ are related by Lorentz transformations;  $\lambda$ and $\lambda'$ are the length scales associated with light propagation in the corresponding Lorentz frames.
 So, there is only one option to construct an expression for energy-momentum  propagation along, or associated with, a null geodesic and that is to look into the scales  involved in the expression of the phase. Doing so, one finds $c$ (Lorentz invariant) and $\lambda$ (observer dependent) are the only two such independent quantities. Therefore, it is  necessary, on dimensional grounds, to introduce a fundamental angular momentum scale ($h$) and write    
\begin{eqnarray}
E=\frac{hc}{\lambda}.
\end{eqnarray}
Hence, one can conclude that association of $h$  is necessary to make sense of  energy-momentum propagation along a null geodesic.  Therefore, if one  considers $h$ to  be the manifestation of quantum physics (which is in accord with the present general understanding), then one encounters a classical-quantum dilemma in special relativity, or rather in the geometric description of spacetime itself. 

 {\it Realizing mass scale variation with a modified thought process:}  To explain the effect of gravity on the rest mass scale of an elementary particle, I shall slightly modify and refine Einstein's thought process.  
  Since $h$ and $c$ are fundamental scales for an elementary particle, following de Broglie, I consider that its rest frame is associated with an energy scale $E_0$ and a frequency scale $(\nu_0)$ by the following relation: $E_0=m_0c^2=h\nu_0$ \cite{confusion,debroglie}. Then, the description goes as follows:
\begin{enumerate}
	\item An observer $O_1$ considers $h$ and $c$ to be given scales and define all other scales in terms of those two. $O_1$ makes measurements with these scales.
	\item $O_1$ studies the decay of a massive elementary particle in its rest frame, e.g. the decay of a neutral pion to two photons: $\pi^0\to \gamma+\gamma$, and measures the amount of released energy in the form of photons with the derived scales.  
	\item Let, $O_1$ measures $\alpha_1$ units of energy with the derived scale $E_1$, i.e. $\alpha_1E_1$ amount of energy is released in the form of radiation as measured by $O_1$. For $O_1$, it implies, $\alpha_1$ amount of mass in the scale $m_1=E_1/c^2$ is the rest mass of the pion, which follows from four momentum conservation. Another observer $O_2$ notes down the whole procedure while relatively at rest with $O_1$.
	
	\item  Then $O_2$ goes to a frame which differs by a gravitational potential from that of $O_1$. $O_1$ repeats the same  experiment and sends the resulting photons from the pion decay to $O_2$.
	
	\item  $O_2$ receives the photons and  measured the energy  with the predefined scales and found that he did not receive $\alpha_1E_1$, but an amount $\alpha_2E_1$.
	
	\item $O_2$ finds that the energy measurement yields $\alpha_1$ if the scale is redefined to be $E_2$ which is related to $E_1$ by the eq.(\ref{e1e2}).
	
	\item $O_2$ concludes that the rest mass of the pion is $\alpha_2m_1$ or $\alpha_1m_2$ where 
	\begin{eqnarray}
	m_2=m_1\(1+\frac{\Phi}{c^2}\)\label{mvary}.
	\end{eqnarray}  
	This implies, if $O_2$ studies the decay of a pion by bringing it to relatively at rest with respect to him/her, then the rest mass will come out to be $\alpha_1$ if measured with the scale $m_2$, but not $m_1$.
	 \end{enumerate}


Although it sufficed for the above thought process to consider only the kinematics (the four momentum conservation), the study of the dynamics of a $\pi^0$ decay suggests that the fine structure constant also needs to be considered as a fundamental constant alongside $h$ and $c$ \cite{peskin}. I may emphasize that  the ratios of rest mass scales associated with different elementary particles, remain the {\it same}, individually for $O_1$ and $O_2$. 

 To mention, exactly like the gravitational redshift has been discussed in terms of geodesics in \cite{wald}, one can think of a similar process here. For example, imagine two space stations $S_1$ and $S_2$ following two different timelike  geodesics. 
 $O_1$ performs the experiment in $S_1$ and transmits the radiation through a window towards $S_2$. $O_2$ receives the radiation through a window at $S_2$. Both $O_1$ and $O_2$ consider $h$ and $c$ to be given.  Rest of the steps of the procedure remain alike.



{\it Compton scattering and effect of gravity:} Now, let me discuss a table-top experiment, that can be performed in two  frames that differ by a gravitational potential. Consider the Compton effect i.e. scattering of x-ray by an electron \cite{compton}. If x-ray, with an associated length scale, $\lambda_0$, is incident on an electron, then it is scattered at an angle $\theta$ with the incident direction. The scattered x-ray is associated with an increased length scale $\lambda_{\theta} (>\lambda_0)$. The increment in the length scale of the x-ray is given by
\begin{eqnarray}
	\Delta \lambda:=\lambda_{\theta}-\lambda_0=\lambda_e(1-\cos\theta).
\end{eqnarray}
where $\lambda_e:=h/m_e c$ is {\it a} length scale associated with the electron, known as Compton length scale (see \cite{remark} for a remark), $m_e$ is the rest mass scale associated with  the electron. Now, consider the {\it same} Compton scattering experiment is performed by $O_1$ and $O_2$. By `same', I mean, $O_1$ and $O_2$ incident same $\lambda_0$, in their respective frames, on an electron and study the same  $\lambda_\theta$. Now, according to eq.(\ref{mvary}), the rest mass of an electron are different for $O_1$ and $O_2$. Therefore, one has the following result:
\begin{eqnarray}
\Delta \lambda= \lambda_1\(1-\cos\theta_1\)=\lambda_2\(1-\cos\theta_2\)\label{dlm}
\end{eqnarray}
where $\lambda_1=\frac{h}{m_1 c},~~~ \lambda_2=\frac{h}{m_2 c}$ are the Compton length scales of the electron and $m_1,m_2$ are the rest masses of the electron for $O_1$ and $O_2$ respectively. Combining eq.(\ref{mvary}) and eq.(\ref{dlm}), one has the following result
\begin{eqnarray}
&&	\frac{m_2}{m_1}=\frac{1-\cos\theta_2}{1-\cos\theta_1}=\(1+\frac{\Phi}{c^2}\)\\
\implies && \frac{\Phi}{c^2}=\frac{(\cos \theta_1-\cos\theta_2)}{(1-\cos\theta_1)}\label{exact}.
\end{eqnarray}

{\it In a Schwarzschild spacetime:} 
Now, let me bring in the spacetime language and set the observers and radiation propagation along geodesics, which I have already mentioned earlier. 
Let me consider $O_1$ and $O_2$ freely falling along two timelike geodesics at different constant values of the radial coordinate  of a Schwarzschild spacetime \cite{wald}; $O_1$ and $O_2$ have radial coordinates $r_1$ and $r_2$ respectively, with $r_2=r_1+H$ and $H>0$. Then, one has the following approximate expression:  
\begin{eqnarray}
\frac{\Phi}{c^2}
\simeq r_0\[ \frac{1}{r_1+H}-\frac{1}{r_1} \]=\[ \frac{1}{x+\Delta x}-\frac{1}{x} \]\simeq -\frac{\Delta x}{x^2},~~
\label{e8}
\end{eqnarray}  
where $r_0=\frac{GM}{c^2}$, $M$ is the mass scale associated with the Schwarzschild spacetime \cite{remark2},  $r_1=x r_0, H=(\Delta x)r_0, x\gg 1,~ \Delta x > 1$ and the last step of eq.(\ref{e8}) holds for $\Delta x\ll x$.  
Now, let $\theta_2=\theta_1+\delta$, where $\delta$ is a  function of $\Delta x$. Then, the right hand side of eq.(\ref{exact}) can be approximated, up to the leading order in $\delta$, to obtain
\begin{eqnarray}
 \frac{(\cos \theta_1-\cos\theta_2)}{(1-\cos\theta_1)}\simeq \delta \cot \frac{\theta_1}{2}.\label{e9}
\end{eqnarray}
Therefore, using the results of  eq.(\ref{e8}) and eq.(\ref{e9}), back in eq.(\ref{exact}), one obtains the result
\begin{eqnarray}
\delta\simeq -\frac{\Delta x}{x^2}  \tan\frac{\theta_1}{2}.
\end{eqnarray}

Just as an example, if one considers $O_1$ to be at earth's radius ($6370$ km \cite{earth}) and $O_2$ on the MICROSCOPE satellite, which is further $700$ km radially outward  \cite{microscope}, then ${\Delta x}/{x^2}\simeq 700/(6370)^2\simeq 1.72\times 10^{-5}$. Then, 
\begin{eqnarray}
\delta_{MICRO}\simeq -1.72\times 10^{-5}\tan\frac{\theta_E}{2}.
\end{eqnarray}
Here, $\theta_E$ is the deflection angle of the scattered x-ray photon on earth's surface and $\delta_{MICRO}$ is the change in the deflection angle while the experiment is performed on the MICROSCOPE.

{\it Conclusion:}  Planck constant is necessary to describe energy-momentum flow along a single null geodesic and therefore, it is intrinsically associated with spacetime geometry. It has {\it a priori} nothing to do with the notion of `quantum'. Further, the rest mass scale variation  is a hitherto unexplored facet of Einstein's equivalence principle. The suggested experiment is first of its kind as it reveals the effect of gravity on the Compton length scale of an elementary particle. A verification will only be suggestive of the fact that what one understands by `kilogram' on earth, is different than the notion of `kilogram' on the moon.

\section*{Acknowledgement}
I am grateful to Benito Juarez Aubry and Srijit Bhattacherjee for several critical discussions regarding this work. This work is supported by the Department of Science and Technology of India through the  INSPIRE Faculty Fellowship, Grant no.- IFA18-PH208.


\begin{thebibliography}{77}
\bibitem{scale} Let me clarify why I associate the word `scale' with all the physical dimensions in the whole discussion. The quantities like mass, length, time, etc. are measured with some predefined scales set by the experimenter. Otherwise, within the theory, one can not assign values to the parameters  that appear in the equations. For example, when one says that ``the rest mass of a particle $m_0=3$ kilograms'' it implies that the person has already set a predefined notion of `1 kilogram'.  Without such a predefined notion, $m_0$ is just a symbol representing  an intrinsic `scale' associated with the particle, which is interpreted as the physical dimension of `mass'.
\bibitem{restjust} I use the word `rest' before `mass scale' so as to emphasize its uniqueness due to Lorentz invariance. The frame dependent mass scale of a point particle is always greater than unity measured with its unique rest mass scale, as is seen in special relativity.
\bibitem{comment1}  $\Delta\nu_{cs}$\cite{cs1} is now used as a standard to define other units \cite{standards}. Using this standard, variation of larger time scales due to  gravitational effect was verified by Hafele and Keating \cite{hk1,hk2}. Optical clocks may take the place of atomic clocks in future \cite{opclock} and this might put the claim of variation of Caesium standards due to gravitational potential \cite{bl} on a sound footing. It is interesting to note that the corresponding Caesium standard length scale  $\sim 32.612 ~\text{mm}$~
	 and the optical standard length scale $\sim 10 $mm are both large compared to the Schwarzschild length scale of the earth $r_{E}=GM_E/c^2\sim 4.437~ \text{mm}$. Therefore, even with the optical clocks it will not be possible to verify the effect of gravitation on scale variation below the background scale set by the earth's gravity. 
\bibitem{equivalence} In this article `equivalence principle'  implies Einstein's version of the principle that was introduced in ref.\cite{einequi}. See \cite{will,liberati} and the references there in to get a detailed account on different types of equivalence principles.
\bibitem{point} The notion of a {\it point} was defined properly by Euclid as follows: {\it A point is that which has no part}. (p.153 of  ref.\cite{elements2}). Therefore, when the behaviour of an object is studied as a whole, it is  considered as a point and called a {\it point particle}  \cite{landau}.
\bibitem{interesting} It is interesting, as well as ironical, because Einstein's own paper concerning the photoelectric effect \cite{einphoto} and Planck's paper on black body radiation spectrum \cite{planck1,planck2}, all of which contained the idea of light quanta, leading to the proportionality of energy and frequency, were published before ref.\cite{einequi} appeared in the literature. The proportionality constant is now known as Planck's constant. 
\bibitem{remark} There is another length scale associated with a mass scale $m$ and it is $Gm/c^2$ where $G$ is the gravitational constant.  This is not used here. This type of length scale involving $G$ can not be associated with radiation. 
\bibitem{remark2} In the Schwarzschild metric there are two mass scales involved. The one which appears in the metric itself, generally denoted by $M$, is called the `mass' associated with the spacetime. It is a scale that appears in the solution of the Einstein equation as an integration constant \cite{schw}. The other one is the mass scale associated with the test particles, say $m_{test}$,  which are used as probes to study the time-like geodesics. It is a general understanding that one must have $m_{test}\ll M$ so that the metric remains undistorted (e.g. see \cite{liberati}).
\end{thebibliography}
\end{document}